\documentstyle[epsfig,12pt]{article} 

\begin{document} 
\topmargin 0pt 
\oddsidemargin 5mm

\setcounter{page}{1}
\hspace{8cm}{Preprint YerPhI 1483(20)-96} \\
\vspace{2cm}
\begin{center}
{GENERATOR-INVERTOR-DAMPER SYSTEM OF ELECTRON (POSITRON)} \\
{BUNCHES MOVING IN COLD PLASMA FOR DEVELOPMENT OF STRONG }\\
{ACCELERATING ELECTRIC FIELD}\\
{\large A.Ts. Amatuni, S.G. Arutunian, M.R. Mailian}\\
\vspace{1cm}
{\em Yerevan Physics Institute}\\
{Alikhanian Brothers' St. 2, Yerevan 375036, Republic of Armenia}
\end{center}
overdense plasmas

\vspace {5mm}
\centerline{{\bf{Abstract}}}

It is shown that high accelerating gradient can be obtained in a
specially constructed system of electron (positron) bunches, moving in cold 
plasma with definite density. These combined bunch systems do not generate 
the wake fields behind them and can pass through the plasma column in a 
periodic sequence. The consideration is carried out numericaly and analyticaly
in one dimensional approach, (which can be applied to finite system when 
its transverse dimensions are larger than plasma wave length, divided by 
$2\pi$).
The possibilities of the experimental tests by measuring the predicted energy
gain are discussed on the examples of Argonne Wakefield Accelerator and
induction linac with typical parameters.

\section{\small{INTRODUCTION}}

Langmuir already noticed, that the beam, which has passed the plasma 
column, contains a significant portion of electrons with the energies higher
than the initial energy of the beam. In the recent times various groups 
(see e.g. \cite{A}-\cite{C}) also observed experimentally the effect of the
selfacceleration.

The possibility to accelerate the electrons of the bunches, moving in 
plasma, was considered numericaly in \cite{D}, \cite{E} and some attempts 
of the analytical treatment of the problem have been performed in \cite{F}-
\cite{I}.

In the work \cite{J}, which is based on the previous results \cite{K} used
as a zero order approximation, nonlinear dynamics of the one-dimensional non
rigid ultrarelativistic bunch of electrons, moving in the cold plasma, is
considered by multiple scales perturbative approach. Square root of the 
inverse Lorentz factor of the bunch electrons is taken as a small parameter.

It was shown \cite{J}, in particular, that for specialy constructed system
consisted from two bunches, first one dense enough $n_b^{(1)} \gg
\frac{1}{2} n_0$, and the second one with low density $n_b^{(1)} \ll
\frac{1}{2}n_0$, the electrons on the rare side of the second bunch, where
electric field $E$ is negative and large, can be significantly
accelerated. In the first approximation of multiple scales method, the
change of the momentum of the bunches electrons is given by $\triangle
p_b=-eEt$, where $t$ is a acceleration (deacceleration) time interval
which, due to applicability of the multiple scales method, is $t \leq
\omega_p^{-1}\gamma^{1/2}$. 

The essential restriction on the acceleration rate in the considered cases
comes from the steady state assumption adopted in \cite{J}. If plasma 
electrons momenta in the wake field behind the first dense relativistic 
bunch, moving in underdense plasma, exciedes same critical value, allowed by 
steady state assumption, the plasma electron density behind the bunch turn 
out to be negative, which probably means that wake wave breaking limit is 
achieved. Restriction on the plasma electrons momenta inside the bunch 
automatically restricts the maximum value of the generated electric field.

In the present work another proper combination of the charged particles 
$(e^{\pm})$ bunches is proposed, which is constructed in such a way, that 
it does not generate the wake fields behind the combined bunch at all. The 
plasma electrons momenta, inside the combined bunch being negative, remains
in the allowed domain and their maximum value is restricted only by the 
total momentum conservation low. The maximum of the subsequent electric 
field then also can be large enough.

The combined bunch in the considered case consists (at least) of three
parts:first one is a relativistic negative particles bunch, with density 
$n_b^{(1)} \gg \frac{1}{2}n_0$, which generates in a plasma inside the bunch 
strong electric field, which brakes the bunch particles (generator). 
The second bunch is a low density bunch with particles $n_b^{(2)}
<\frac{1}{2}n_0$ (it can be negative particles bunch, positive particles 
bunch), which inverts 
the sign of the electric field generated by  the first bunch (invertor). 
The third bunch with the parameters, which can coincide with that of the 
first one, provided on its rear part the zero values of electric field and
plasma electrons momenta, coincided with the same values on the front
of the first bunch (damper). It means that plasma behind the system of 
bunches remains unperturbed and no wake field is generated.

Acceleration of the particles of the rear part of the second and of the 
third bunches, where electric field is negative, will take place. The goal 
of the following calculations is to estimate the maximum possible values 
of the acceleration parameter.

In the next section of the work the vivid approach to the problem, based 
on the numerical solutions of the subsequent nonlinear equations and 
their graphic representations on phase plane are presented. 
The third section devoted to the analytical treatment of the problem,
 which based on the exact solutions of the equations and  provides 
the general expressions for the energy gain, acceleration rate and 
acceleration length, the lengths of the different parts of the combined 
bunch as a functions of charge densities and Lorentz-factor of the 
bunches. Finally, the numerical examples are presented, which consider the 
possibilities of the experimental observations of the predicted 
selfacceleration of charged particles, in particular, on Argonne Wakefield
Accelerator (AWA) and induction linacs.

\section{\small{PHASE PLANE IMAGES OF THE COMBINED BUNCHES, MOVING IN COLD
PLASMA}}

Consider a plane rigid electron bunch with the infinite transverse 
dimensions and length $[0,d]$ along the $z$-axis. Velocity of the bunch 
in the lab system is $v_z=v_0 \quad (\beta=\frac{v_0}{c})$, plasma is cold
with the immobile ions. The charge density of the plasma electrons is 
$n_b$;the case, when $n_b > \frac{1}{2}n_0$ called underdense regime, $n_b <
\frac{1}{2}n_0$-overdense regime. The plasma electrons motion equations
described in hydrodynamic approximation is
\begin{equation}
\label{A}
\frac{\partial \rho_e}{\partial t'}+\beta_e\frac{\partial \rho_e}{\partial z'}
=-E,\rm{or} \quad (\beta_e-\beta_0)\frac{d\rho_e}{d\tilde{z}}=-E;
\end{equation}
Continuity eq. for plasma electrons is 
\begin{equation}
\label{B}
\frac{\partial n_e}{\partial t'}+\frac{\partial}{\partial z'}
(\beta_e {n'}_e)=0,\rm{or} \quad \frac{d}{d\tilde z}\left[{n'}_e
(\beta_e-\beta)\right]=0,
\end{equation}
and Maxwell (Coulomb) equation for the electric field is
\begin{equation}
\label{C}
\frac{dE'}{dz}=\frac{E'}{d{\tilde{z}}'}=n'_0-n'_e-n'_b
\end{equation}

In eqs. (\ref{A}-\ref{C}) $$t'=\omega_pt \quad 
z'=k_pz,\omega_p=\left(\frac{4\pi e^2n_b}{m}\right)^{1/2},$$ $$E=(4\pi 
nmc^2)^{1/2}E'=\frac{mc\omega_p}{e}E',n'_e=\frac{n_e}{n},n'_b=\frac{n_b}{n}$$
$$\rho_e=\frac{\rho_{ez}}{mc},\beta_e=\frac{v_{ez}}{c},\gamma=(1-\beta^2)^
{-1/2};$$
$\tilde{z}=z-v_0t$ (steady state approximation, $n$- is an arbitrary 
density, which finaly will be chosen).

Continuity eq. (\ref{B}) has an integral
\begin{equation}
\label{D}
n_e(\beta-\beta_e)=n_0\beta;
\end{equation}
constant is defined from the boundary condition, $z=d,\beta_e=0,n_e=n_0$. 
In (\ref{D}) and in what follows the superscripts "prime" are omitted.

From eq. (\ref{A}) it follows that 
\begin{equation}
\label{E}
E=-\frac{d\Phi}{d\tilde{z}}\equiv -\Phi';\Phi \equiv 
\sqrt{1+\rho_e^2}-\beta\rho_e 
\end{equation}
and eqs (\ref{B})-(\ref{C}) can be written in the form
\begin{equation}
\label{F}
\Phi''-\beta\gamma^2\frac{\Phi}{(\Phi^2-\gamma^{-2})^{1/2}}=\alpha-\gamma^2
\end{equation}
(see also \cite{K}, \cite{M}-\cite{N}).

Eq. (\ref{F}) has an "energy" integral
\begin{equation}
\label{G}
{\cal{E}}=\frac{1}{2}\Phi'^2+\gamma^2\left[\Phi-\beta(\Phi^2-\gamma^{-2})
\right]-\alpha \Phi
\end{equation}
where $\alpha \equiv  n_b/n_0$;the boundary conditions at $\tilde{z}=d$
are $\rho_e=0,\Phi=1,\Phi'=0$ and gives ${\cal{E}}=1-\alpha$, but in what 
follows the eq. (\ref{G}) considered at the arbitrary boundary conditions.

The integral (\ref{G}) allowed to interpret equations (\ref{A}-\ref{C})
as  equations for the point with unit  mass with the "coordinate" $\Phi$
and "velocity" $\Phi'$ moving in the potential $U=U_0+U_1=
\gamma^2\left[\Phi-
\beta\left(\Phi^2-\gamma^{-2}\right)^{1/2}\right]-\alpha 
\Phi$.

On Fig. 1 the $U$ as a function of $\Phi$ is drawn for different values of
$\alpha$ and for $\gamma=10$. The negative values of $\alpha=\frac{n_b}{n_0}$
correspond to the bunch of positive charged particles, positive $\alpha$
corresponds to the bunch of negative charged particles. For $\alpha<0,
\gamma^{-1} \leq \Phi \leq 1$ and $\Phi \ge 1$for $\alpha >0$, when boundary 
conditions, when
$\bar{z}=d$, are $\Phi=1$ and $\Phi'=0$. From Fig. 1 it is seen that for 
$\alpha <0$
the motion is always periodic, it is also periodic for $0 \leq \alpha \leq 
1/2$, and non periodic for $\alpha >1/2$ \cite{K}.
\epsfig{file=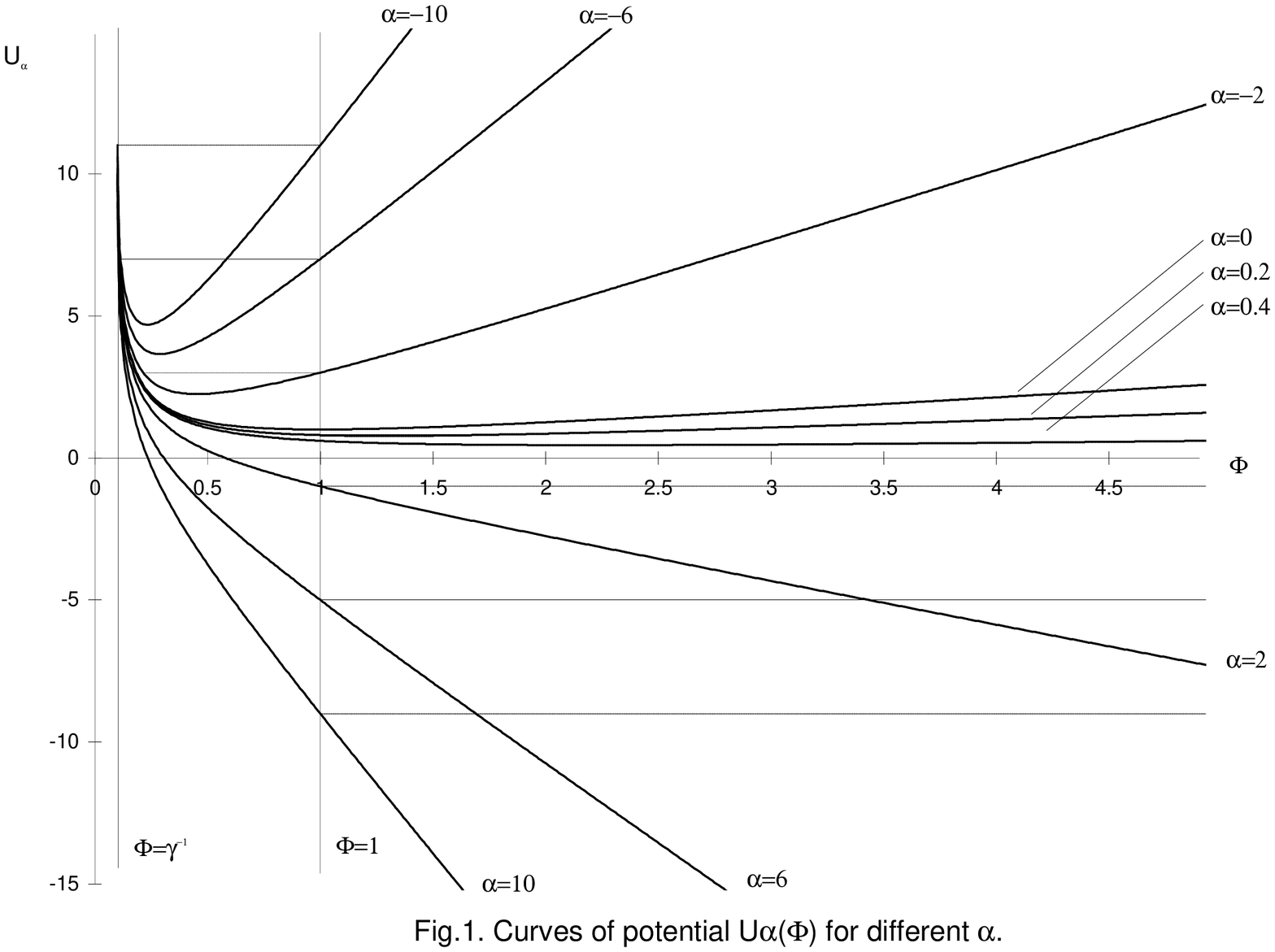, height=10cm, width=13cm, angle=270}
\epsfig{file=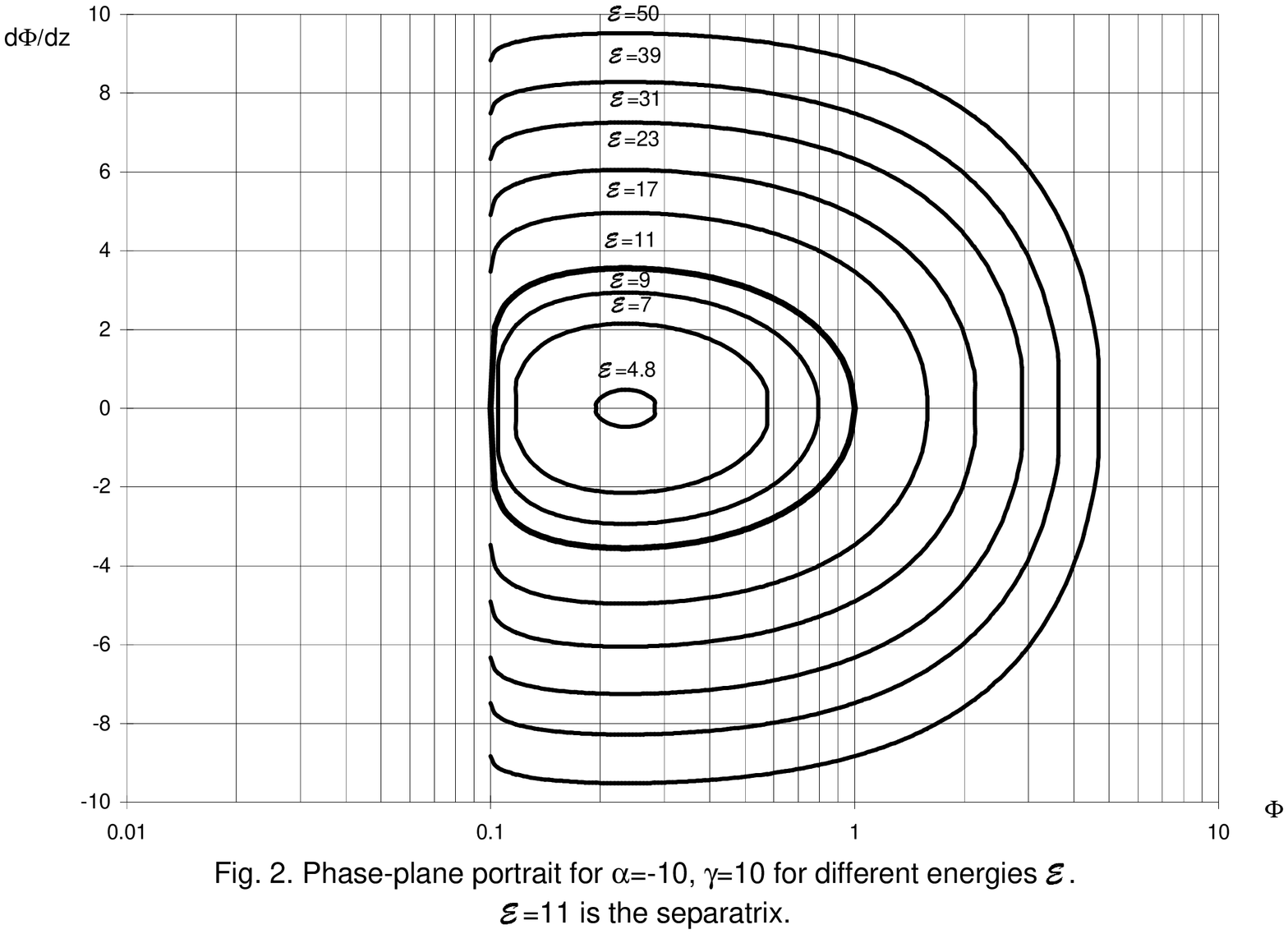,height=8cm,width=10cm,angle=270}
\epsfig{file=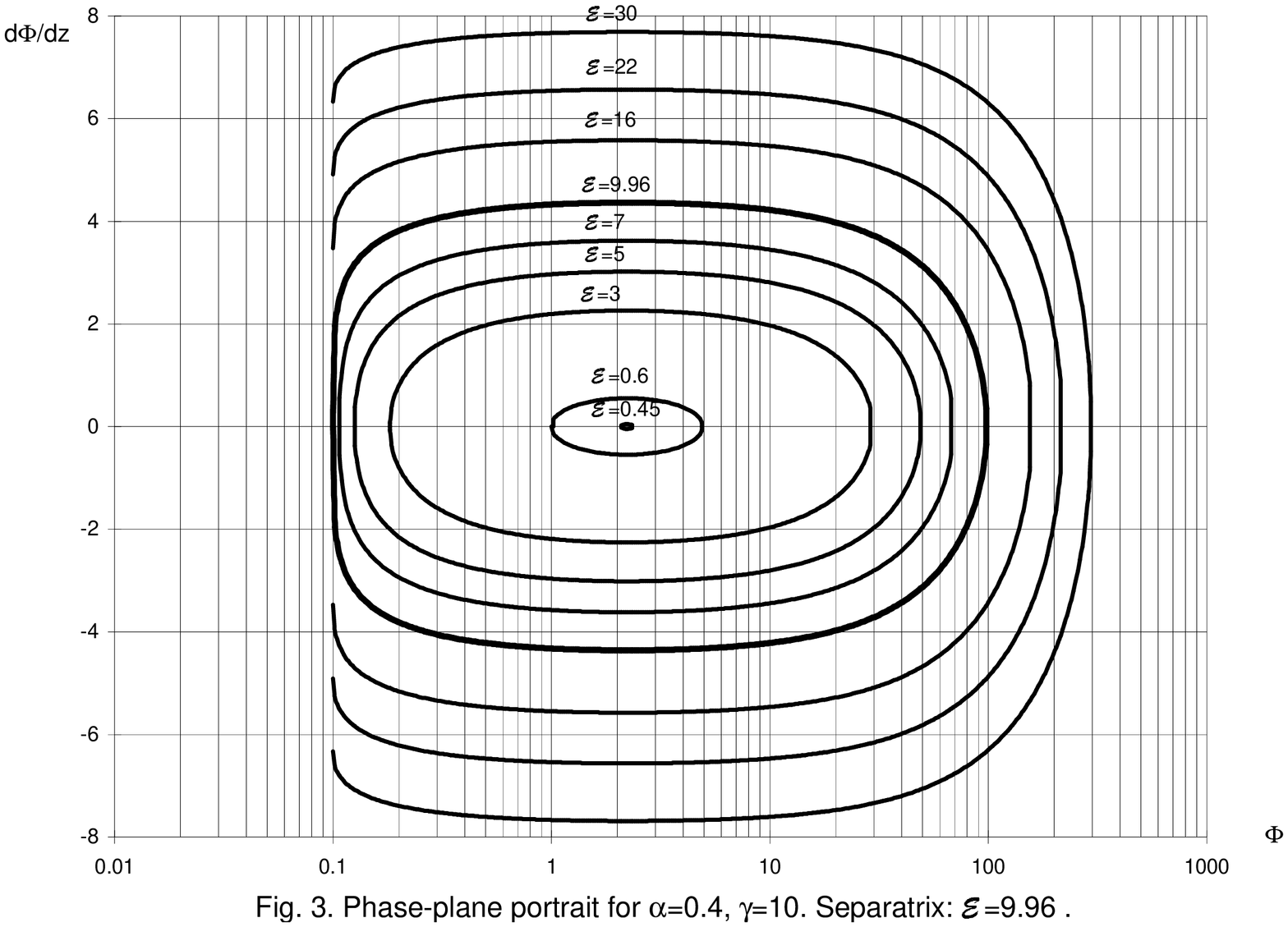,height=8cm,width=10cm,angle=270}
\epsfig{file=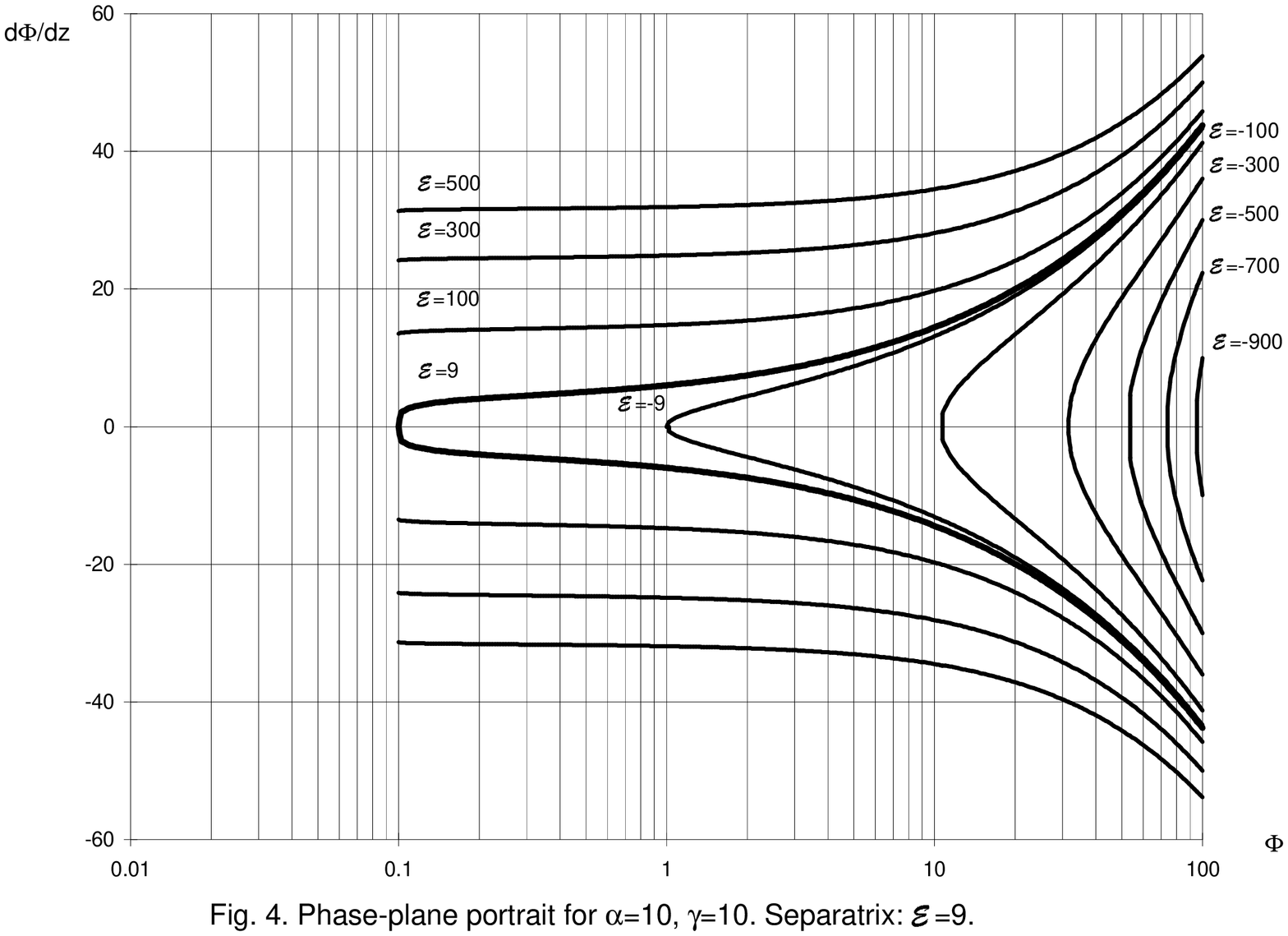,height=8cm,width=10cm,angle=270} 
Fig. 2, 3, 4 represent the phase plane $(\Phi',\Phi)$ portrait of the 
system of cold
plasma-rigid charged particle bunch in the cases, when $\alpha=-10,0.4,10$
and $\gamma=10$ at different boundary conditions (different ${\cal{E}}$). 
The closed phase trajectories correspond to periodic motion.

Trajectories on Fig. 2, 3, 4 can serve as a building blocks to construct the
phase portraits of the combined bunches, moving in cold plasma. It must be
represented by a closed loop, starts from the boundary condition 
$\Phi=1,\Phi'=0$
and accomplished at the same point $\Phi=1,\Phi'=0$ on the phase plane. 
An example of such a loop is presented on Fig. 5.
The loop on Fig. 5b
from the point $0: \quad \Phi=1,\Phi'=0 \quad (\rho_e=0,v_e=0,E=0$, when 
$\tilde{z}
=d$) and the curve $OA$ represents the motion of the bunch of negative
charged particles (generator) $\alpha^{(1)}=10$, when at the rear side of 
the bunch very strong but positive electric field $E_0=-\Phi'$ can  be 
generated. Taking this point as a new boundary conditions for the second 
bunch of positive charged particles (invertor), the trajectory is turned to
the point $B$, where electric field is zero and then to the point $A'$
where the field $-E_0$ is obtained. From point $A'$ to the point $0$ the
trajectory described the motion of the third bunch of the negatively
charged particles (damper) on the rear side of which the initial boundary
conditions exist. It means that no wake field excites and the plasma 
behind the third bunch remains unperturbed. 
As a consequence, the above mentioned restriction on the value of the field
$E_0$, which arouse from the condition of the stability of the wake 
field, is removed.
\epsfig{file=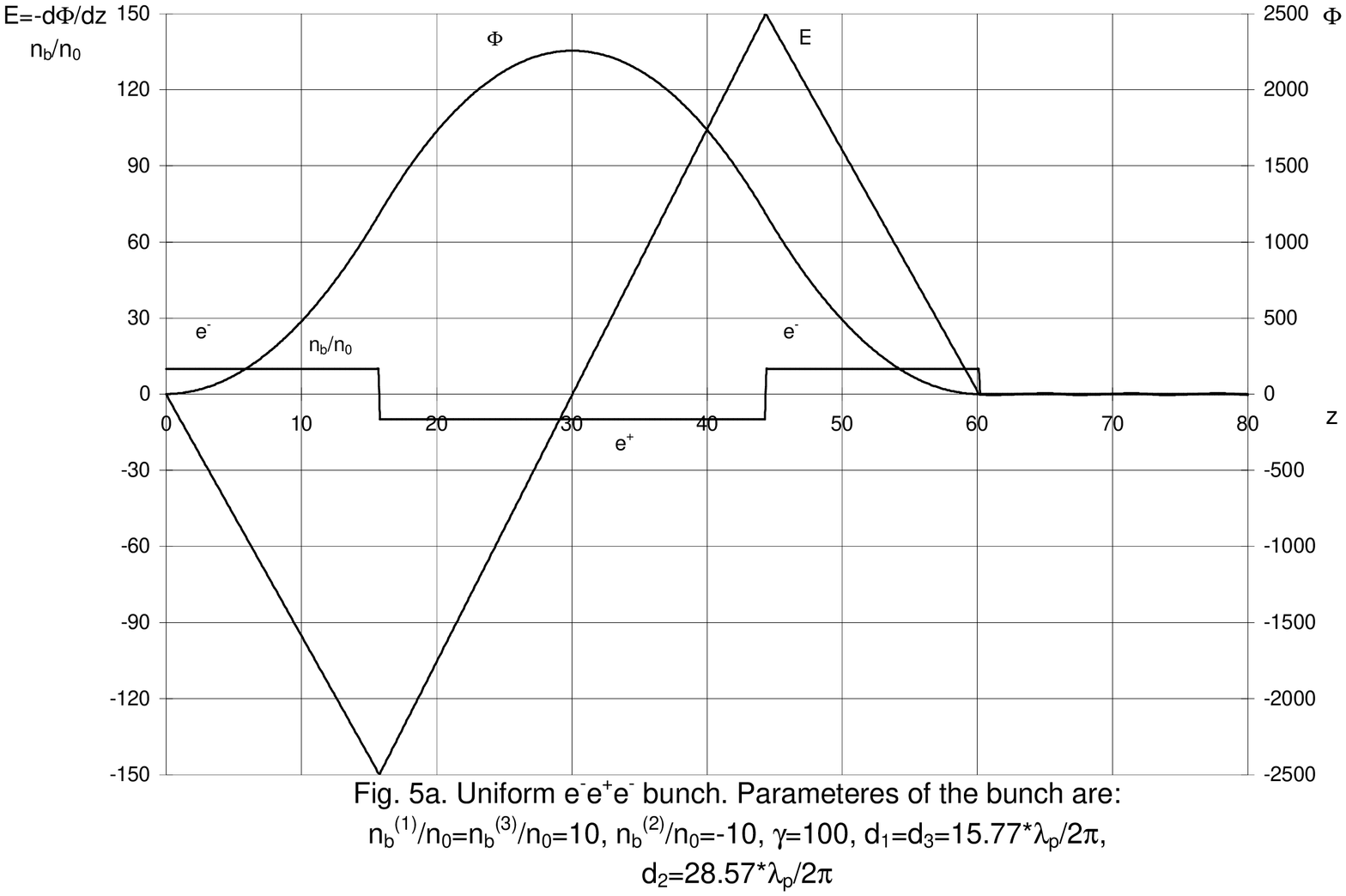,height=7cm,width=8cm,angle=270}
\epsfig{file=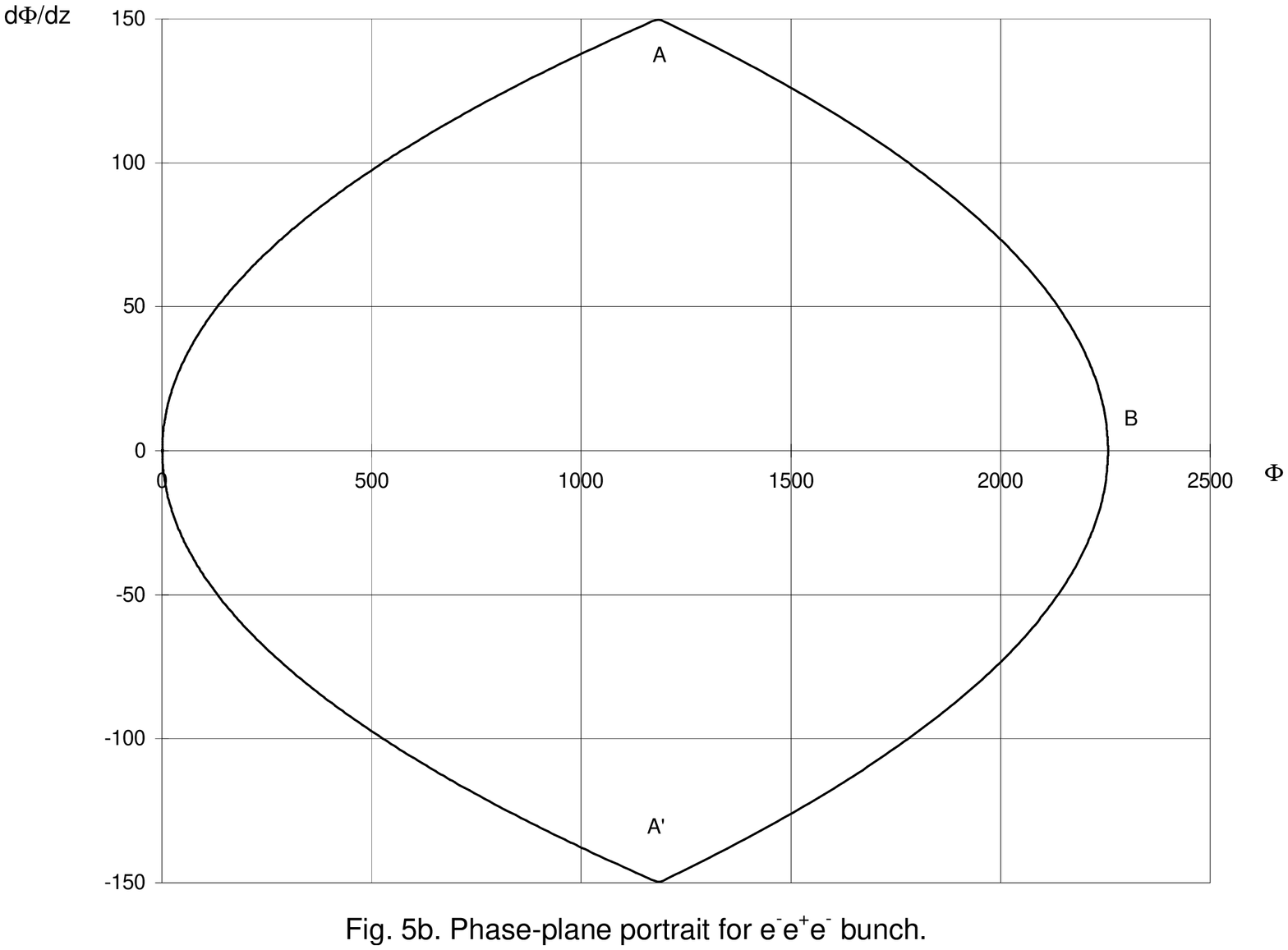,height=6cm,width=8cm,angle=270}

This is not only the condition for the applicability of steady state regime
for the description of the zero order approximation in the multiple scales
approach, but also open the possibility to repeat more than once the 
procedure, sending the next combined bunches into the unperturbed by 
previous bunches plasma.

The Lorentz factor of all three bunches in the presented case is, of 
course, the same $\gamma=100$, and $$\alpha^{(2)}=-\frac{n_b^{(2)}}{n_0}=-10,
\alpha^{(3)}=\alpha^{(1)}=\frac{n_b^{(3)}}{n_0}=10.$$ The total length
of the positron bunch is $28.57\frac{\lambda_p}{2\pi}$ and lengths of the 
first and third electron bunches are $\sim 15.77\frac{\lambda_p}{2\pi}$
each.

On Fig. 5a the potential $\Phi$ and electric field $E=-\frac{\partial \Phi}
{\partial \tilde{z}}$ as a function of $\tilde{z}$ are presented. It is 
evident from Fig. 5a that positrons of the first part of the second 
bunch, where $E>0$, can be accelerated as well as all  electrons of the third
bunch $(E<0)$.

The similar picture may be obtained by construction of the combination 
of three bunches with negatively charged particles (electrons). First bunch
is dense $n_b^{(1)}/n_0 \gg 1$, second has a low density $\frac{1}{2}n_0-
n_b^{(2)} \gg \gamma^{-2}$, and third is again dense $n_b^{(3)} \gg n_0$.
\epsfig{file=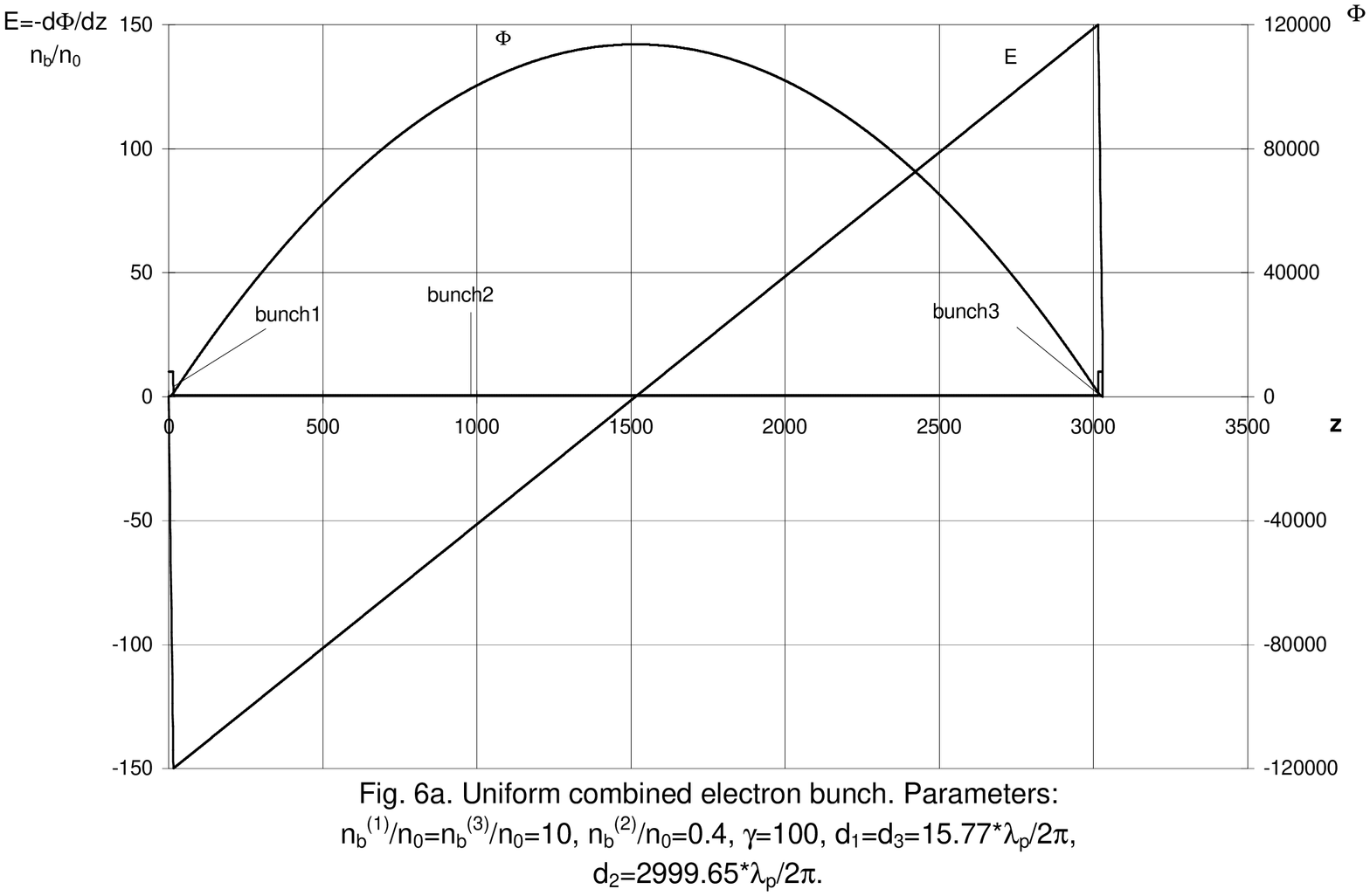,height=8cm,width=8cm,angle=270}
\epsfig{file=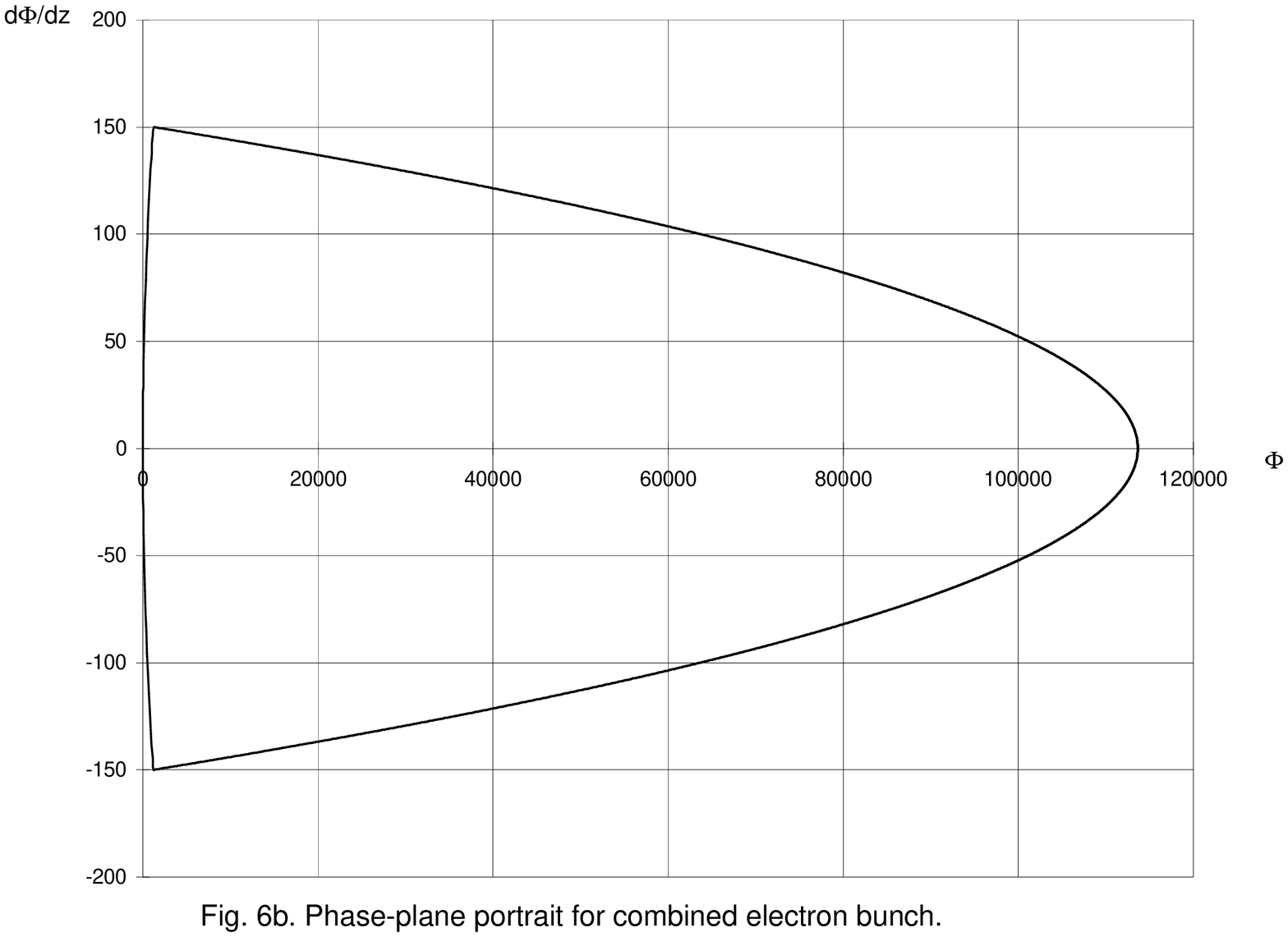,height=8cm,width=8cm,angle=270}
The
example of such a combination is presented on Fig. 6a, 6b, where electric
field and phase portrait are drawn for the case, when $$n_b^{(1)}=n_b^{(3)}=
10n_0,n_b^{(2)}=0.4n_0,\gamma=100.$$ 
It is seen from Fig. 6a, that the 
length $d_2$ of the second bunch-invertor is too large in considered 
case. Acceleration of the bunches electrons take place when $E<0$ i.e. mostly
in the rear side of the second bunch-invertor and, at some extend, in the 
third bunch-damper.

The role of the bunch-invertor may play also the plasma itself 
$(n_b^{(2)}=0)$. In such a case the electric field interacts on plasma 
electrons, which are moving in the direction opposite to that of 
bunch, with the initial velocity equal to zero. So they will accelerate in 
the region where $E>0$, and drag in the region where $E<0$. The electrons 
of the third bunch, where $E>0$ will accelerate. 
The charge distribution inside the bunches can be nonuniform.

On Figs. 7a, 7b
the results of the numerical solution of eqs. (\ref{A}-\ref{C}) are 
displayed for the case of sinusoidal charge distribution inside the electron-
positron-electron bunches span by plasma in-between. The bunches have the 
equal lengths $d=15\frac{\lambda_p}{2\pi}$, devided by plasma columns with 
lengths $d_0=60\frac{\lambda_p}{2\pi},\gamma=100$, $$\frac{n_b^{(1)}}{n_0}=
\frac{n_b^{(3)}}{n_0}=10,\frac{n_b^{(2)}}{n_0}=-11.48.$$
\epsfig{file=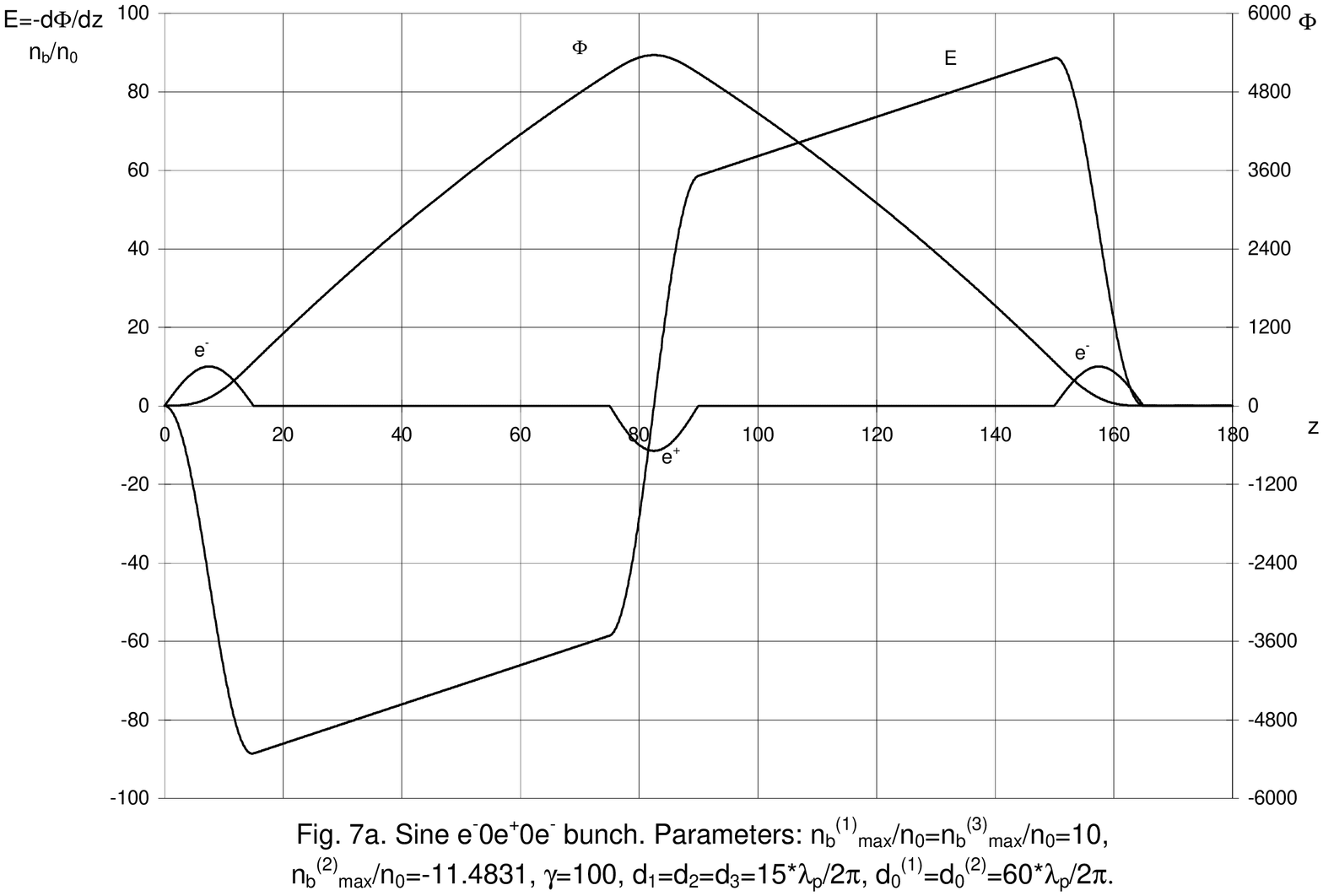,height=8cm,width=8cm,angle=270}
\epsfig{file=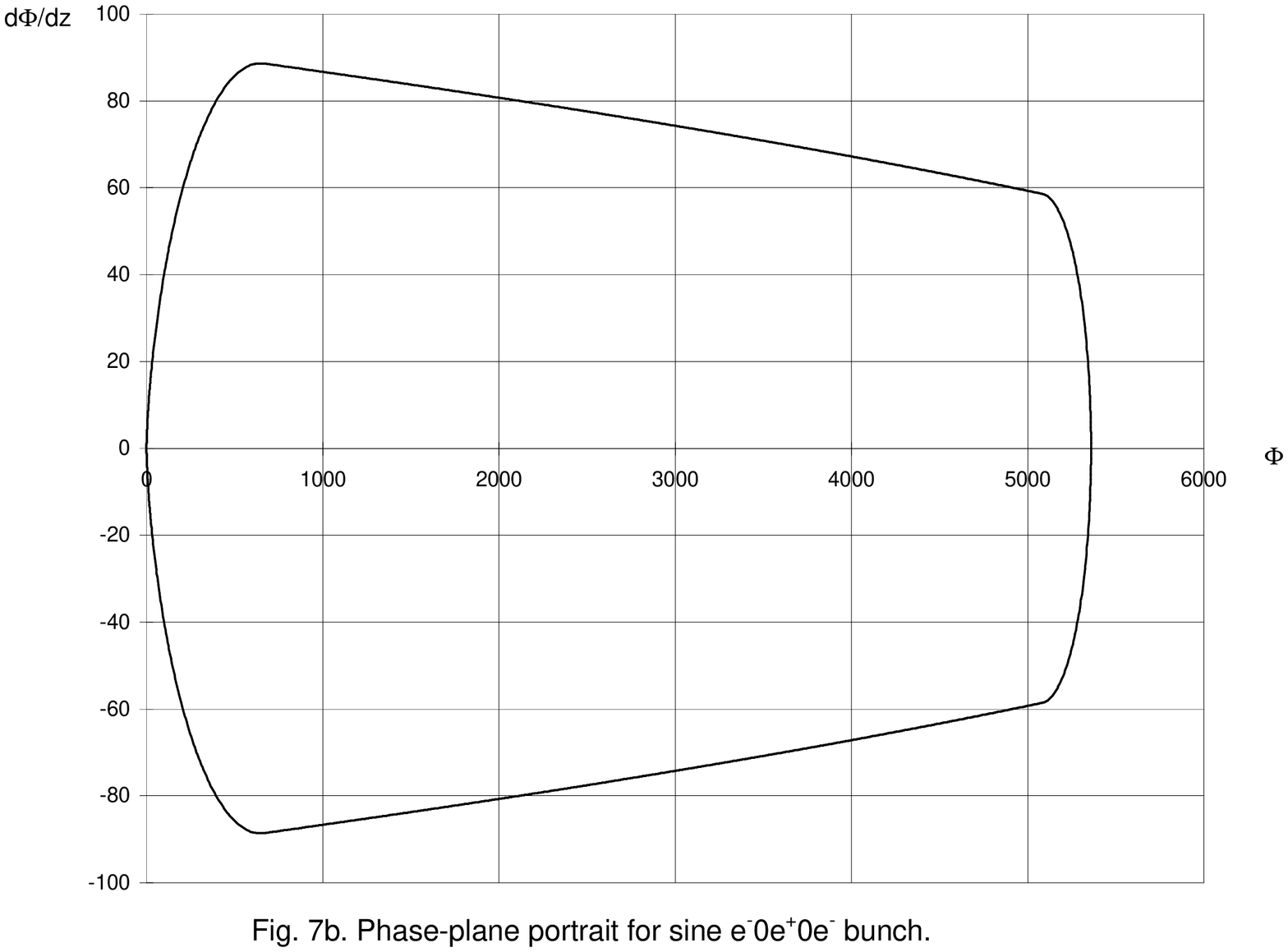,height=8cm,width=8cm,angle=270}
It is neccessary to note that the wake field behind the combined bunch due to 
small inaccuracies, presented in actual calculations
practically exists, but the amplitude of this wake is small and far from 
wave breacking limit. This last condition defines the tolerances on the 
parameters of combined bunch and must be taken into account in actual 
calculations for the proposed experimental tests. In the case of the 
sequence of the combined bunches, passing through plasma, the 
conditions, which
define the tolerances on bunch parameters, will be more severe.

\section{\small{ANALYTICAL APPROACH}}

In this section the results of the exact analytical solutions of eqs.  
(\ref{A}-\ref{C}) for the combined bunch with the uniform charge 
distributions inside the constituting bunches are presented. Results
for single rigid (electron) bunch with the zero boundary conditions 
$\rho_e=0,E=0,$ when $\tilde{z}=d$) are obtained in \cite{K}.

Boundary conditions at the end of the first electron bunch moving in 
underdense plasma $n_b \gg \frac{1}{2}n_0$ are $\rho_e=\rho_0,E=E_0$, where
\begin{eqnarray}
\label{H}
E_0=&\pm\left[2\left(n_b^{(1)}-n_0\right)\right]^{1/2}\left\{\left(1+\rho_0^2
\right)^{1/2}-1-\alpha^{(1)}\beta\rho_0\right\}^{1/2} \\ \nonumber
\alpha^{(1)} \equiv & n_b^{(1)}/n_b^{(1)}-n_0
\end{eqnarray}

The field inside the second electron bunch, moving in overdense regime 
$n_b^{(2)}<1/2n_0$ with the same velocity is 
\begin{eqnarray}
\label{I}
E^{(2)}=&\pm\left[2\left(n_0-n_b^{(2)}\right)\right]^{1/2}\left\{a-\left(
1+\rho_e^2\right)-\alpha^{(2)}\beta\rho_e\right\}^{1/2},\nonumber \\ 
\alpha^{(2)}=&\frac{n_b^{(2)}}{n_0-n_b^{(2)}}<1,\quad \alpha^{(2)}\beta<1, 
\nonumber \\
a \equiv &
1-\frac{n_b^{(1)}-n_b^{(2)}}{n_0-n_b^{(2)}}\left\{\left[1-(1+\rho_0^2)^{1/2}
\right]-\beta\rho_0\right\};
\end{eqnarray}
When $\rho_0 \rightarrow 0 \quad a \rightarrow 1$, when $$\rho_0<0,|\rho_0|
\gg 1$$
\begin{equation}
\label{J}
a \approx \frac{n_b^{(1)}-n_b^{(2)}}{n_0-n_b{(2)}}(1+\beta)|\rho_0| \gg 1
\end{equation}

When plasma electron momenta is equal to 
\begin{equation}
\label{K}
\rho_{\pm}=-\frac{a\alpha^{(2)}\beta}{1-\left(\alpha^{(2 )}\beta\right)^2}
\pm \left[\frac{\left(a\alpha^{(2)}\beta\right)^2}{1-\left(\alpha^{(2)}
\beta\right)^2}+\frac{a^2-1}{1-\alpha^{(2)}\beta}\right]^{1/2}
\end{equation}
the $E^{(2)}=0$;when $|\rho_0| \gg 1$ and $a \gg 1$
\begin{equation}
\label{L}
\rho_{-} = -\frac{a}{1-\alpha^{(2)}\beta},|\rho_{-}|>|\rho_0|
\end{equation}
Note that it is impossible to obtain $E=0$ and $\rho_e=0$ simultaneously
using only the second (electron, $n_b^{(2)}>0$, or positron, $n_b^{(2)}>0$) 
bunch.

The second root $\rho_+$will not be reached in the considered cases.

The electric field $E^{(2)}$ has a maximum at 
\begin{equation}
\label{M}
\rho_e^{max}=-\frac{\alpha^{(2)}\beta}{\left[1-(\alpha^{(2)}\beta)^2\right]
^{1/2}}
\end{equation}
In the considered cases $\rho_e^{max}$ is choosen according the condition
$$|\rho_-|>|\rho_0|>\left|\rho_e^{max}\right|.$$

The length of the second bunch is choosen in such a way that at the end 
of it $E^{(2)}=-E_0$, and $\rho_e=\rho_0$ (see Fig. 6).

The third bunch then can be similar to the first one in order to dump 
the electric field and plasma electron momenta to their initial zero 
values, in order to dump wake fields completely. For this end the length 
$d^{(2)}$ must be found, integrating the equation
\begin{equation}
\label{N}
dz=\frac{\pm[\beta(1+\rho_e^2)-\rho_e] d\rho_e}
{[2(n_0-n_b^{(2)})]^{1/2}(1+\rho_e^2)^{1/2}[a-(1+\rho_e^2)^{1/2}-\alpha^{(2)}
\beta\rho_e]^{1/2}}
\end{equation}
by two steps-from $\rho_0$ to $\rho_-$ and from $\rho_-$ to $\rho_0$, taking
sign $+$ when $\rho_e$ increases with the $\tilde{z}$ and sign $-$, when
$\rho_e$ decreases with the $z$. In both cases $|\rho_e| \geq \rho_0 \gg 1$
and the integration simplifies essentially, when $(1+\rho_e^2)^{1/2}$ 
replaced by $|\rho_e|$;the result of integration, using (\ref{J}, 
\ref{L}), is
\begin{equation}
\label{O}
d_2^e=\frac{4(1+\beta)|\rho_0|^{1/2}}{[2(n_0-n_b^{(2)})]^{1/2}(1-(\alpha^{(2)}
\beta))^{1/2}}\left[\frac{(n_b^{(1)}-n_b^{(2)})}{(n_0-n_b^{(2)})}
\frac{(1+\beta)}{(1-\alpha^{(2)}\beta)}-1\right]^{1/2}
\end{equation}

In the case, when the second bunch is consisted from positive particles
(positrons) the length of the second bunch, obtained by the similar 
calculations, is
\begin{equation}
\label{P}
d_2^p=\frac{4(1+\beta)|\rho_0|^{1/2}}{[2(n_0+{\bar{n}}_b^{(2)})]^{1/2}(1+
{\bar{\alpha}}^{(2)}\beta)^{1/2}}\left[\frac{(n_b^{(1)}+{\bar{n}}_b^{(2)})}
{(n_0+{\bar{n}}_b^{(2)})}
\frac{(1+\beta)}{(1+{\bar{\alpha}}^{(2)}\beta)}-1\right]^{1/2}
\end{equation}
where ${\bar{\alpha}}^{(2)} \equiv 
\frac{{\bar{n}}_b^{(2)}}{n_0+{\bar{n}}_b^{(2)}}$

In the positron case $n_b^{(2)}$ can be choosen as 
${\bar{n}}_b^{(2)}=n_b^{(1)}
\gg n_0$ for the effective inversion of the electric field $E_0$. Comparison
of the eqs. (\ref{O}) and (\ref{P}) for $$n_b^{(2)} \ll n_0,n^{(1)} \gg n_0
\quad \bar{n}_b^{(2)}=n_0,\beta =1,$$ shows, that in this case
\begin{equation}
\label{Q}
\frac{d_2^p}{d_2^e}=\frac{n_0}{2n_b^{(1)}} \ll 1
\end{equation}
i.e. the dense ${\bar{n}}_b^{(2)} = n_0$ positron bunch-invertor would be 
much shorter than electron bunch with $n_b^{(2)} <1/2n_0$. The length of the
first (and third) electron bunch can be obtained similiarly or from the 
results of work \cite{L} for the case, when 
$$1\ll\frac{n_b^{(1)}}{n_0}\ll2\gamma^2,|\rho_0| \gg 1,\beta \rightarrow 1,$$
and is
\begin{equation}
\label{R}
d_1=\beta|\rho_0|^{1/2} =d_3
\end{equation}

The possible largest plasma electron momenta inside the first bunch in 
the considered case differs from that found in \cite{K} by wake wave 
breaking limit and can be estimated from momentum conservation (see also 
\cite{J}),
\begin{equation}
\label{S}
n_b^{(1)}\beta\gamma d^{(1)} \ge \int^{d_1}_0 \rho_e(\tilde{z})n_e(\tilde{z})
dz=\frac{n_0|\rho_0|d_1}{4}
\end{equation}

For the estimate in (\ref{S}) the expression for $n_e$, obtained from 
continuity equation $$n_e(\tilde{z})=\frac{n_0\beta(1+\rho_e)^{1/2}}
{\beta(1+\rho_e^2)^{1/2}-\rho_e},$$
is used, which approximated by linear one in interval from $n_e=n_0$ at 
$p_e=0$ up to the value $n_e=\frac{n_0\beta}{1+\beta}$ for large $\rho_e$.

From (\ref{S})
\begin{equation}
\label{T}
1 \ll |\rho_0| \le \frac{4n_b^{(1)}}{n_0}\gamma,
\end{equation}
and corresponding electric field from (\ref{H}) and (\ref{T}) is
\begin{equation}
\label{U}
|E_0|=2\left(n_b^{(1)}\right)^{1/2}|\rho_0|^{1/2} \le 
\frac{4n_b^{(1)}}{n_0^{1/2}}\gamma^{1/2}
\end{equation}

As it was shown in \cite{J} the change of the bunch electron momenta 
$\triangle p$, in the first approximation, which can be found from the 
second order equations, is 
\begin{equation}
\label{V}
\triangle p=p_{b1}=-E(\tilde{z})\tau
\end{equation}
where $E(\tilde{z})$ is the electric field in the zero order 
approximation inside the rigid bunch and $\tau=\omega_pt \le 
\gamma^{1/2}$. The limit on acceleration time interval follows from the domain
of validity of the applied multiple scales approach.

The length of the plasma column needed for the change of the momenta 
given by (\ref{V}) is, in ordinary units
\begin{equation}
\label{W}
l=ct \le \frac{c}{\omega_p}\gamma^{1/2}=\frac{\lambda_p}{2\pi}\gamma^{1/2}
\end{equation}

The acceleration rate (gradient), in ordinary units is
\begin{equation}
\label{X}
G=\frac{c\triangle p}{el}
\end{equation}

\section{\small{POSSIBILITIES OF THE EXPERIMENTAL TESTS}}

The formulae obtained in sec. 3 for the parameters of proposed device in
ordinary units and $n=n_0$ are from (\ref{T}):
\begin{equation}
\label{Y}
|\rho_0| \le \frac{4n_b^{(1)}}{n_0}\gamma mc,
\end{equation}
from (\ref{U}) 
\begin{equation}
\label{Z}
E_0 \le 4\left(\frac{n_b^{(1)}}{n_0}\right)\gamma^{1/2}\frac{mc\omega_p}{e},
\end{equation}
from (\ref{V}), (\ref{Z})
\begin{eqnarray}
\label{AA}
\triangle p_b&=&eE_0t \le 
4\frac{n_b^{(1)}}{n_0}mc\gamma=\frac{4n_b^{(1)}}{n_0} p_b; \\ \nonumber
\frac{\triangle p_b}{p_b}& \le & 4\left(\frac{n_b^{(1)}}{n_0}\right);\quad
t \leq \omega_p^{-1}\gamma^{1/2}
\end{eqnarray}.

The accelerating gradient from (\ref{X}), (\ref{AA}) is
\begin{eqnarray}
\label{AB}
G=&\frac{c\triangle p_b}{el}=\frac{\triangle p}{et} \sim
4\left(\frac{n_b^{(1)}}{n_0}\right)\frac{mc\gamma^{1/2}\omega_p}{e}\\ 
\nonumber
G \sim &\frac{4\pi}{\lambda_p}\left(\frac{n_b^{(1)}}{n_0}\right)\gamma^{1/2}
MV/cm, \\ \nonumber
\lambda_p=&\frac{2\pi c}{\omega_p}=6.08\cdot10^{-4}\left(\frac{3\cdot 
10^{19}}{n_0}\right)^{1/2}cm
\end{eqnarray}

The acceleration length (\ref{V}):
$$ l=ct\le\frac{\lambda_p}{2\pi}\gamma^{1/2}.$$

The lengths of the first and third electron bunches (\ref{R}), (\ref{Y}) are:

\begin{equation}
\label{AC}
d_1=d_3 \le 
\frac{\lambda_p}{\pi}\gamma^{1/2}\left(\frac{n_b^{(1)}}{n_0}\right)^{1/2}
\end{equation}

The length of the second electron bunch (\ref{O}), (\ref{Y}) is
\begin{eqnarray}
\label{AD}
d_2^e \le &\frac{8\lambda_p}{\pi}\gamma^{1/2}\left(\frac{n_b^{(1)}}{n_0}
\right) \\ \nonumber
n_b^{(2)} \ll & n_0 \ll n_b^{(1)}
\end{eqnarray}

The length of the second positron bunch (\ref{P}), (\ref{Y}) is:
\begin{eqnarray}
\label{AE}
d_2^p \le &\frac{4\lambda_p}{\pi}\gamma^{1/2} \\ \nonumber
{\bar n}_b^{(2)} =& n_b^{(1)} \gg n_0;
\end{eqnarray}

According to (\ref{Z}) it is possible to obtain essential acceleration 
fields when
the previously accelerated dense bunches of negative charged particles 
(electrons) are used. The lengths of such a bunches, according to 
(\ref{AC}), imcreases as $\sim \gamma^{1/2}$, so it seems that they must 
be the bunches from high current accelerators with the energies up to tens of 
Mev. Even at these energies maximal acceleration gradient is high 
enough, it can exceed the acceleration gradient of the ordinary linacs by 
a several orders of magnitude.

Consider the experimental possibilities of proof-of-principle experiment
on existing accelerators or test facilities. It seems that Argonne 
Wakefield Accelerators (AWA), which accelerates very dense bunches, is 
suitable for above mentioned proposes.

AWA \cite{Q} is $20 Me\rm{V}$ electron accelerator which accelerate the 
bunches of $20 p$ sec  full width (the length $d_{1exp}=0.6cm$), diameter
about $0.1cm$ \cite{R}, the total charge of the bunch 100nC, i.e. 
$n_b^{(1)}=2\cdot 10^{14}cm^{-3}$. If $n_0$ is choosen as $n_0=\frac{1}{5}
n_b^{(1)}=4\cdot 10^{13}cm^{-3}$, then the optimal length of the first
bunch from (\ref{AC}) is $d_1=2.4cm$, i.e. 4 times larger than the 
experimentally existing one $d_{1exp}=0.6cm$.

The role of the invertor can serve the plasma column itself. Witness 
bunch, which has 10ps duration, less than 1nC total charge, and energy 
4Mev can be 
used as accelerated bunch, being placed somewhere between first and third
bunches. Third bunch (damper) has similiar parameters as the first one 
and must be placed at distance $d_2=8\left(\frac{n_b^{(1)}}{n_0}\right)^{1/2}
d_1 \approx 43cm$ from the end of the first bunch. Experimentally it is
possible to put it on the distance $d_{2exp}=46cm$, because the frequency 
of the RF-cavities is $\nu=1.3 GHz \quad (d_{2exp}=nc/\nu=n\cdot 
23cm, n=1,2,3,\cdots)$. If the length of the first bunch could be made 4 
times large than existing one and total charge of the bunch will be also 4
times larger-400nC, then the change of the energy of the witness bunch 
will be, according to (\ref{AA}) $$\triangle {\cal E}_w \le 4\frac{n_b^{(1)}}
{n_0}{\cal E}_1=400Me\rm{V} \gg {\cal E}_w=4Me\rm{V}.$$

For the existing experimental parameters of AWA:$d_1=0,5cm,d_2=23cm$, total
charge of bunch 70nC, $n_b^{(1)}=1.4\cdot 10^{14}cm^{-3}$ results of the 
numerical calculations, along the lines presented in Sec. 2, are shown on 
figures 8a and 8b.
The role of bunch-invertor plays the plasma column, with density $n_0$ as a
free parameter, which was been chosen from the necessity to invert the 
field, produced by the first bunch, in such a way, that the succesive 
bunch, with the same parameters as the first one, moving at fixed (by 
RF-frequency) distance from the first one, $d_2=23cm$, will be able to dump 
the field to zero. 
The sought density of plasma $n_0$ was found equal to $n_0 \approx 1.17\cdot
10^{13}cm^{-3},\lambda_p=0.973cm,n_b/n_0=11.95$. Then the maximum value 
of the accelerated electric field is equal to $E_{max} \approx 100Mv/cm$, and
$E<0$ on the rear part of the second (third) bunch. So about the half of the 
electrons of the second (third) bunch, as it can be seen from Fig. 8a, 
can be accelerated during the time $t \sim \omega_p^{-1}\gamma^{1/2}$ up to 
energy ${\cal{E}}_f \approx 70 Me\rm{V}$.
\epsfig{file=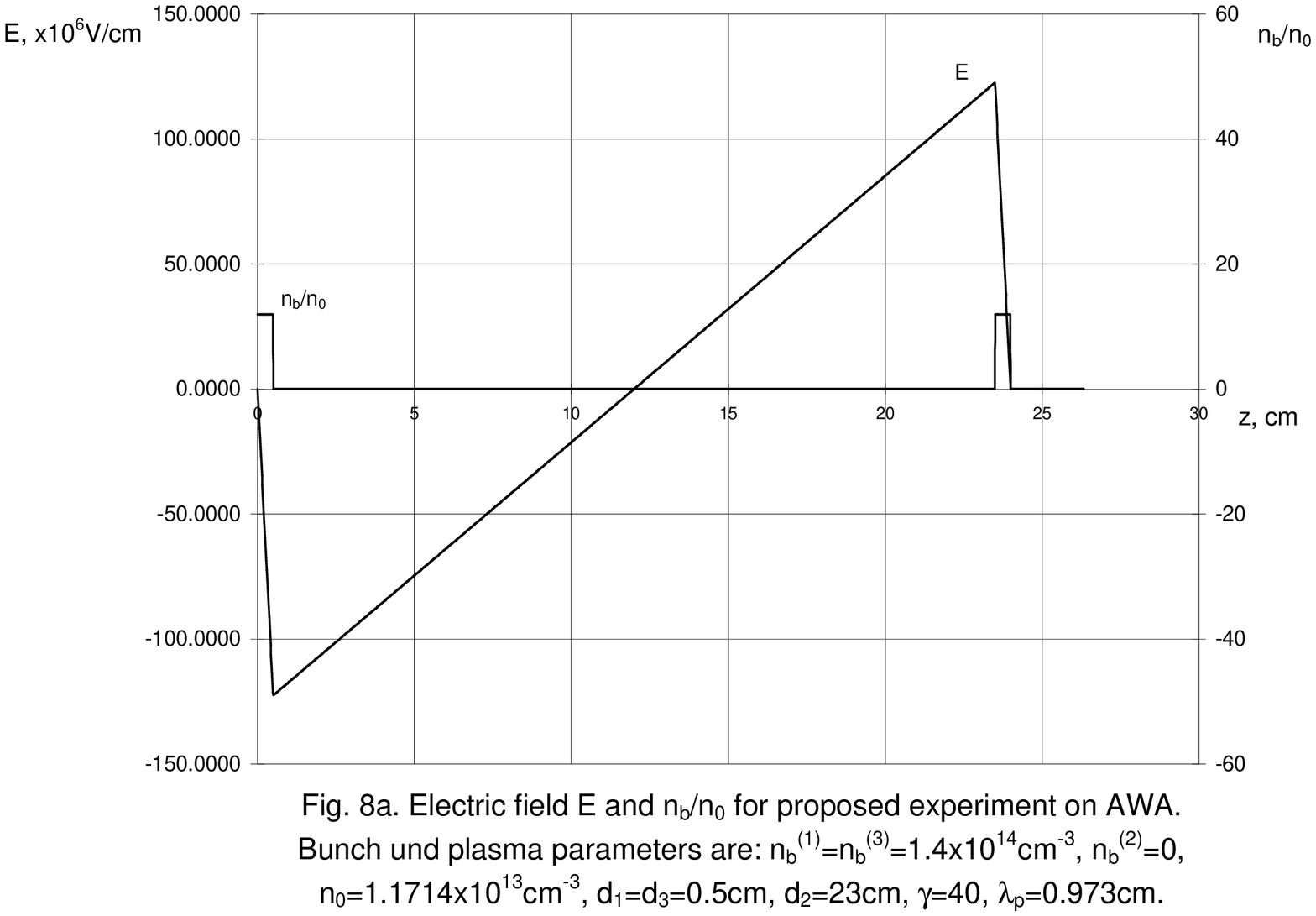,height=8cm,width=8cm,angle=270}
\epsfig{file=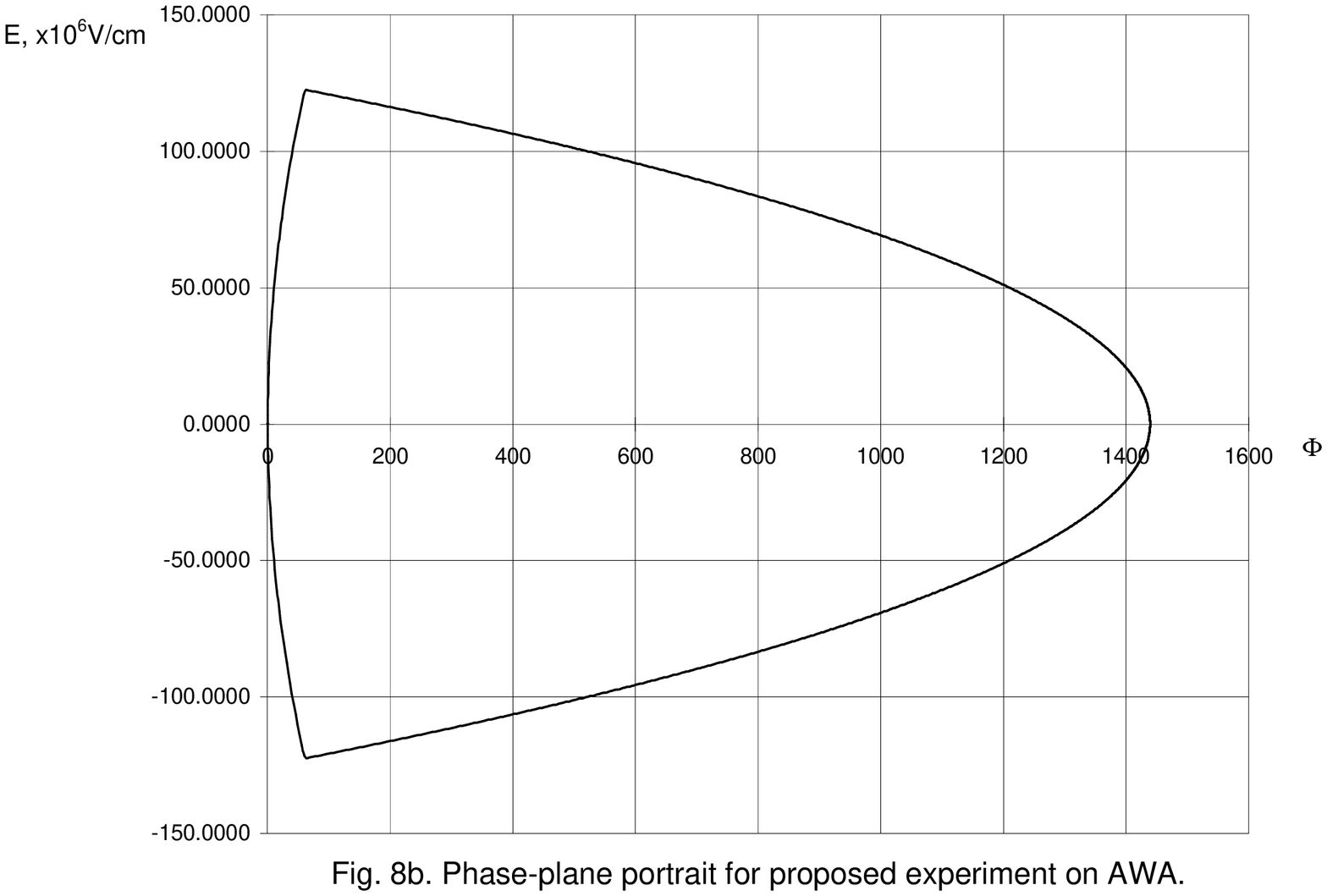,height=8cm,width=8cm,angle=270}
Proposed experiment can serve as a prof-of-principle one;plasma remains 
practically unperturbed after the passage of the first and the second 
bunches, so it is possible to inject the next couple of the bunches and 
also obtain acceleration, again leaving the plasma behind the bunches in 
practically unperturbed state and so on. As it was above mentioned in 
order to obtain the optimal experimental conditions, it is necessary to
enlarge the micropulse duration, leaving its charge density the same. This 
will rise the maximum obtainable electric field and energy gain. Of 
course, the obtained numbers for $E_{max}$ and ${\cal{E}}$ are at some extent 
tentative;it is neccessary to remember that they are obtained in the model
of the bunch with the infinite transverse dimensions and uniform charge 
distribution. More detailed calculations must be performed in the course of 
preparation of the experiment.

However, taking into account the results of  the numerical calculation for
laser driven wake fields in generalized vortex-free plasma \cite{S}, it is 
possible to adopt the statement of authors \cite{S}, that "even when driver's
width approaches $c/\omega_p$ the 1-D (nonlinear theory) predictions are 
still a good guide for determining the accelerating field strength". In the
considered case of AWA the diameter of the bunch is $D=0.1cm \le \frac{c}
{\omega_p}=\frac{\lambda_p}{2\pi}=0.15cm$, so for the existing parameters of
AWA predictions of 1-D nonlinear theory can be valid.

For the improved parameters of AWA, as it was shown, $n_0=4\cdot10^{13}cm^{-3}$
and $D=0.1cm \ge \frac{c}{\omega_p}=\frac{\lambda_p}{2\pi}=0.08cm$ and 
the situation is more favourable.

For Accelerator Test Facility (ATF) at BNL the situation in the considered
respect is worse, due to the very narrow bunches $r=0.03cm$ and smaller 
bunch charge $Q=2nC$ \cite{T}. At any case, the considered problem needs 
the development of 2-D nonlinear theory of wake field generation by the 
electron (positron) bunches with the finite transverse dimension.

Another possibility to check the predicted effect is based on the use of 
induction linacs.
Consider the example of the accelerator with the parameters, which are 
within the range of the possibilities of the existing or planned induction 
linacs (see e.g. \cite{O}, \cite{P}):
Energy $\qquad \qquad {\cal{E}}=16 Me\rm{V}$\\
Pulse current $\qquad I=5kA$ \\
Pulse duration $\qquad \tau=2.5nsec$\\
Diameter of the bunch $D=2 cm$\\
Then the charge density of the first and third bunches will be $n_b^{(1)}=
1.6\cdot 10^{11}cm^{-3}$, the plasma density can be chosen as 
$n_0=\frac{1}{5}
n_b^{(1)}$ and then, according to (\ref{AC}, $d_1 \approx 75 cm$, the 
distance between 
second and third bunches, according to (\ref{AD}), is $d_2 \approx 13.4m$ 
(the
plasma column plays the role of invertor). The maximum energy gain of the 
witness bunch of much smaller current (e.g. 0.1 I and 
$\tau<2.5nsec$), placed in the proper position between the first and third 
bunches, according to (\ref{AA}), is $$\triangle {\cal{E}}^{max}=4\cdot 
5\cdot 16=320Me\rm{V}.$$
The condition of the applicability of the 1-D nonlinear thery in this 
case is practically fulfiled:$$D=2cm \le \frac{\lambda_p}{2\pi}=2.3cm.$$ If 
the positron bunch can be used as a second invertor bunch 
with the same current I as the first one, then the length $d_2$, according 
to (\ref{AE}), can be essentially reduced-$d_2^p=134.1cm$, and the part of 
positrons from this bunch can be accelerated, up to the energies 
$\approx 300 Me\rm{V}$.

It is evident, that in all cases considered, the length of the plasma 
column must be 2-3 times larger than the total length of the combined 
bunch, due to transient effects.

The increase of the lengths of the bunches with the energy may be 
considered as tolerable one, when the extended astrophysical objects, as 
e.g. supernovae, are considered. Then proposed mechanism can be applied, for 
example, even for the explanation of the origin of the superhigh energy 
cosmic 
rays. This problem (see e.g. \cite{U}) is now the most important one for 
the astrophysics of cosmic rays with the energies 
${\cal{E}}>10^{17}e\rm{V}$, 
when all known mechanisms of cosmic rays acceleration are uneffective 
\cite{V}.

In order to illustrate this possibility consider the following numerical 
example.
Suppose that the flux of electrons with the initial energy ${\cal{E}}_i=
5 \cdot 10^{16}e\rm{V}$ already obtained by some effective and commonly 
adopted 
mechanism. The flux with the density $n_b^{(1)}=1.5 \cdot 10^{14}cm^{-3}$
passes through the electron-proton plasma with the density $n_0=3 \cdot 
10^{11}cm^{-3}$. Then it followed by positron (\ref{B}) and electron 
(\ref{C}) fluxes with the same densities $n_b^{(2)}=n_b^{(3)}=n_b^{(1)}$. 
Then some part of positron and third-electron bunches can be accelerated up to 
final energy 
$${\cal{E}}_f^{max}=4\left(\frac{n_b^{(1)}}{n_0}\right){\cal{E}}_i=10^{20}
e\rm{V}$$
i.e. reaches the atmost limit of the observed cosmic rays. The length of 
the electron fluxes is 140km, the length of the positron bunch is 25km. 
(Let us mention that up to now observed events at such energies are very 
rare). Considered example is only illustrative and aplicability of the
proposed mechanism to the problem of the origin of the superhigh energy 
cosmic rays needs, of course, more careful consideration.

In addition, it is necessary to mention,that the expression for momentum
change (\ref{AA}) $\triangle p=-eEt$, obtained for the first approximation
of the developed multiple scales method,is valid up to fifth approximation:
\begin{equation}
\label{AF}
\frac{p_b}{mc} \equiv \rho_b=\rho_{b0}+\epsilon\rho_{b1}+\epsilon^5\rho_{b5}+
\epsilon^6\rho_{b6} + \cdots,\quad \epsilon=\gamma_0^{-1/2},
\end{equation}
as it follows from the results of \cite{J}. Hence it is possible to 
assume,that eq. (\ref{AA}) is valid for the time intervals larger than 
$\omega_pt \le \epsilon^{-1}=\gamma^{1/2}$ used in the present work for 
estimites. But in order to proof this assumption it is necessary to 
develop more complicated version of multiple scales method, than used in
\cite{J}, where only three scales are introduced: $$\tilde{z}=z-\beta_0t,
\quad \tau_1=\epsilon t,\quad \zeta_1=\epsilon z.$$ Enlargement of the 
validity domain on $t$ needs consideration of the contributions from the next
scales $$\tau_2=\epsilon^2 t,\quad \tau_3=\epsilon^3t,\quad 
\tau_4=\epsilon^4t$$ to the momentum development (\ref{AF}).

\section{\small{AKNOWLEDGMENTS}}

The authors are very thankful to all participants of the YerPhI Seminar on 
"New Methods of Acceleration" for useful discussion and comments.

The work was supported by ISTC and Minatom of RF.

\end{document}